\documentclass[prl,12pt,superscriptaddress]{revtex4}

\usepackage{hyperref}

\usepackage{ulem}
\usepackage{color}

\usepackage{amssymb}
\usepackage{amsmath}
\usepackage{amscd}

\usepackage{graphicx}
\usepackage{setspace}

\usepackage{longtable}
\usepackage{threeparttable}
\usepackage{multirow}
\usepackage{dcolumn}
\usepackage{booktabs}
\usepackage{tabularx}
\usepackage{supertabular}

\begin{document}

\title{Enhanced perpendicular magnetocrystalline anisotropy energy in an artificial magnetic material with bulk spin-momentum coupling}

\author{Abdul-Muizz Pradipto}
\email[Electronic address: ]{a.m.t.pradipto@gmail.com}
\affiliation{Institute for Chemical Research, Kyoto University, Uji, Kyoto 611-0011, Japan}
\affiliation{Department of Physics Engineering, Mie University, Tsu, Mie 514-8507, Japan}

\author{Kay Yakushiji}
\affiliation{National Institute of Advanced Industrial Science and Technology (AIST), Tsukuba, Ibaraki 305-8568, Japan}

\author{Woo Seung Ham}
\affiliation{Institute for Chemical Research, Kyoto University, Uji, Kyoto 611-0011, Japan}

\author{Sanghoon Kim}
\affiliation{Institute for Chemical Research, Kyoto University, Uji, Kyoto 611-0011, Japan}

\author{Yoichi Shiota}
\affiliation{Institute for Chemical Research, Kyoto University, Uji, Kyoto 611-0011, Japan}

\author{Takahiro Moriyama}
\affiliation{Institute for Chemical Research, Kyoto University, Uji, Kyoto 611-0011, Japan}

\author{Kyoung-Whan Kim}
\affiliation{Institute of Physics, Johannes Gutenberg University Mainz, 55099 Mainz, Germany}
\affiliation{Center for Spintronics, Korea Institute of Science and Technology, Seoul 02792, Korea}

\author{Hyun-Woo Lee}
\affiliation{Department of Physics, Pohang University of Science and Technology, Pohang 37673, Korea}

\author{Kohji Nakamura} 
\affiliation{Department of Physics Engineering, Mie University, Tsu, Mie 514-8507, Japan}

\author{Kyung-Jin Lee}
\affiliation{Department of Materials Science and Engineering, Korea University, Seoul 02841, Korea}
\affiliation{KU-KIST Graduate School of Converging Science and Technology, Korea University, Seoul 02841, Korea}

\author{Teruo Ono}
\affiliation{Institute for Chemical Research, Kyoto University, Uji, Kyoto 611-0011, Japan}


\begin{spacing}{1.4}

\begin{abstract}

We systematically investigate the perpendicular magnetocrystalline anisotropy (MCA) in Co$-$Pt/Pd-based multilayers. Our magnetic measurement data shows that the asymmetric Co/Pd/Pt multilayer has a significantly larger perpendicular magnetic anisotropy (PMA) energy compared to the symmetric Co/Pt and Co/Pd multilayer samples. We further support this experiment by first principles calculations on the CoPt$_2$, CoPd$_2$, and CoPtPd, which are composite bulk materials that consist of three atomic layers in a unit cell, Pt/Co/Pt, Pd/Co/Pd, Pt/Co/Pd, respectively. 
By estimating the contribution of bulk spin-momentum coupling to the MCA energy, we show that the CoPtPd multilayer with the symmetry breaking has a significantly larger perpendicular magnetic anisotropy (PMA) energy than the other multilayers that are otherwise similar but lack the symmetry breaking. 
This observation thus provides an evidence of the PMA enhancement due to the structural inversion symmetry breaking and highlights the asymmetric CoPtPd as the first artificial magnetic material with bulk spin-momentum coupling, which opens a new pathway toward the design of materials with strong PMA.

\end{abstract}

\end{spacing}
\begin{spacing}{1.5}

\maketitle


The interplay between magnetism, electronic structure and spin-orbit coupling (SOC) in materials has led to the possibility to utilize both the charge and spin degrees of freedom of electrons for spintronic devices \cite{Wolf:2001,Chappert:2007,Bader:2010,Ohno:2010}. Among the most important effects emerging from the SOC is the magnetocrystalline anisotropy (MCA), resulting in a preferred magnetization direction with respect to the crystallographic structure of materials \cite{Falicov:1990,Daalderop:1992}. For the practical design of devices, perpendicular magnetic anisotropy (PMA) with respect to surface/interface planes is strongly desired \cite{Ikeda:2010,Chiba:2012,Dieny:2017}, as manipulation of the magnetic moment directions and/or the magnetic domain walls can be done more efficiently \cite{Mangin:2006,Sinha:2013}. 
The origin of MCA has initially been attributed to the orbital localization as a result of reduced dimensionality \cite{Neel:1954,Bruno:1989b}. It was also shown that the MCA is strongly related to the SOC of the electronic states near the Fermi level \cite{Nakamura:2009}. Consequently, manipulation of the band structure around the Fermi level provides a natural way to tune the MCA. This can be achieved by the modification of the orbital occupation, for example via the application of an external gate voltage \cite{Maruyama:2009,Niranjan:2010}, or by direct chemical manipulations of the band structure. The latter is commonly done by doping or impurities \cite{Besser:1967,Khan:2017} or by engineering the materials interfaces \cite{Nakamura:2010,Nakamura:2017}. Another milestone in understanding the MCA comes from the proportionality of MCA and the anisotropy of SOC-induced orbital magnetic moment, which was proposed by Bruno \cite{Bruno:1989}, and has been confirmed in different materials \cite{Weller:1994,Weller:1995,Stohr:1999}.

Another mechanism of MCA, which depends on broken inversion symmetry, has been proposed \cite{Barnes:2014,Kim:2016}. In systems with broken inversion symmetry, the SOC becomes odd in the momentum $\vec{k}$ space \cite{Grytsyuk:2016}. The oddness of SOC in the $\vec{k}$ space is visible for instance from the Rashba model of SOC \cite{Bychkov:1984} which has the form $\mathcal{H}_R=\alpha_R\left(\vec{k}\times\hat{z}\right)\cdot\vec{\sigma}$ that depends on linear term of $\vec{k}$. Here the $\alpha_R$ is the Rashba parameter, and $z$ is the direction of inversion-symmetry-breaking-induced potential gradient. 
We note that the broken inversion symmetry results not only in the linear-in-$k$ contribution but also in higher-odd-order contributions. From now on, we use the term ``Rashba" to describe all odd-order-in-$k$ contributions for simplicity. Since the Rashba interaction only develops in non-centrosymmetric systems, the strength of Rashba parameter can provide an indication to the degree of structural inversion symmetric breaking. Although it was originally proposed in nonmagnetic materials \cite{Bychkov:1984,Picozzi:2014,Manchon:2015}, Rashba-type spin splitting was later observed also in magnetic systems, such as on the Gd(0001) grown on an oxide substrate \cite{Krupin:2005}. Recently, the MCA has been analyzed using the SOC model of Rashba \cite{Barnes:2014,Kim:2016} and it was shown that MCA changes with the increasing Rashba parameter strength, i.e. the more asymmetric the system is, the stronger it develops the MCA. Such description is very insightful for the understanding of MCA, however, its verification is still lacking.

\begin{table*}
\begin{center}
\caption{The MCA energy $E_{\rm MCA}$ per unit volume (in meV/{\AA}$^3$) obtained from a self-consistent treatment of the SOC; the $k_y-$dependent $E_{\rm MCA}^{\pm k_y}$ evaluated within half of the Brillouin zone; and the even and odd contributions to the $k-$dependent $E_{\rm MCA}$ (see text).}
\begin{tabular}{lcc*{1}{D{.}{.}{4}}@{}cc*{1}{D{.}{.}{4}}@{}cc*{1}{D{.}{.}{5}}@{}}
\toprule[1pt]
System &&& \multicolumn{1}{c}{CoPt$_2$} &&& \multicolumn{1}{c}{CoPd$_2$} &&& \multicolumn{1}{c}{CoPtPd} \\ 
 \midrule[0.5pt]
\rule{0pt}{2.5ex}$E_{\rm MCA}$ (SCF) &&& 0.028 &&& 0.027 &&& 0.043 \\
\rule{0pt}{2.5ex}$k_y-$dependent $E_{\rm MCA}$: \\
\rule{0pt}{3.0ex}$E_{\rm MCA}^{+k_y}$ &&& 0.015 &&& 0.013 &&& -0.072 \\
\rule{0pt}{3.0ex}$E_{\rm MCA}^{-k_y}$ &&& 0.015 &&& 0.013 &&& 0.113 \\
\rule{0pt}{3.0ex}$E_{\rm MCA}^{\rm even}$ (from Eq. \eqref{even}) &&& 0.030 &&& 0.026 &&& 0.041 \\
\rule{0pt}{3.0ex}$E_{\rm MCA}^{\rm odd}$ (from Eq. \eqref{odd})&&& 0.000 &&& 0.000 &&& 0.185 \\
\bottomrule[1pt]
\end{tabular}
\label{tabSummary}
\end{center}
\end{table*}

We report in this work our analysis on the effects of asymmetric stacking of Co$-$Pt/Pd-based multilayer systems by combining experimental magnetic measurements and first-principles calculation. Co/Pt and Co/Pd-based layered structures have been well known to exhibit strong PMA of about 3$-$10 Merg/cm$^3$ \cite{Mangin:2006,Garcia:1989,Canedy:2000,Meng:2006,Yakushiji:2010,Kato:2012}. We show that the broken inversion symmetry plays a significant role in the appearance of PMA in this material. These results present an evidence to the PMA driven by structural asymmetry as suggested in previous theoretical analyses \cite{Barnes:2014,Kim:2016} and may provide a useful guide in the design of materials with strong PMA. 

\begin{figure}[b]
\begin{center}
\includegraphics[width=1.0\columnwidth]{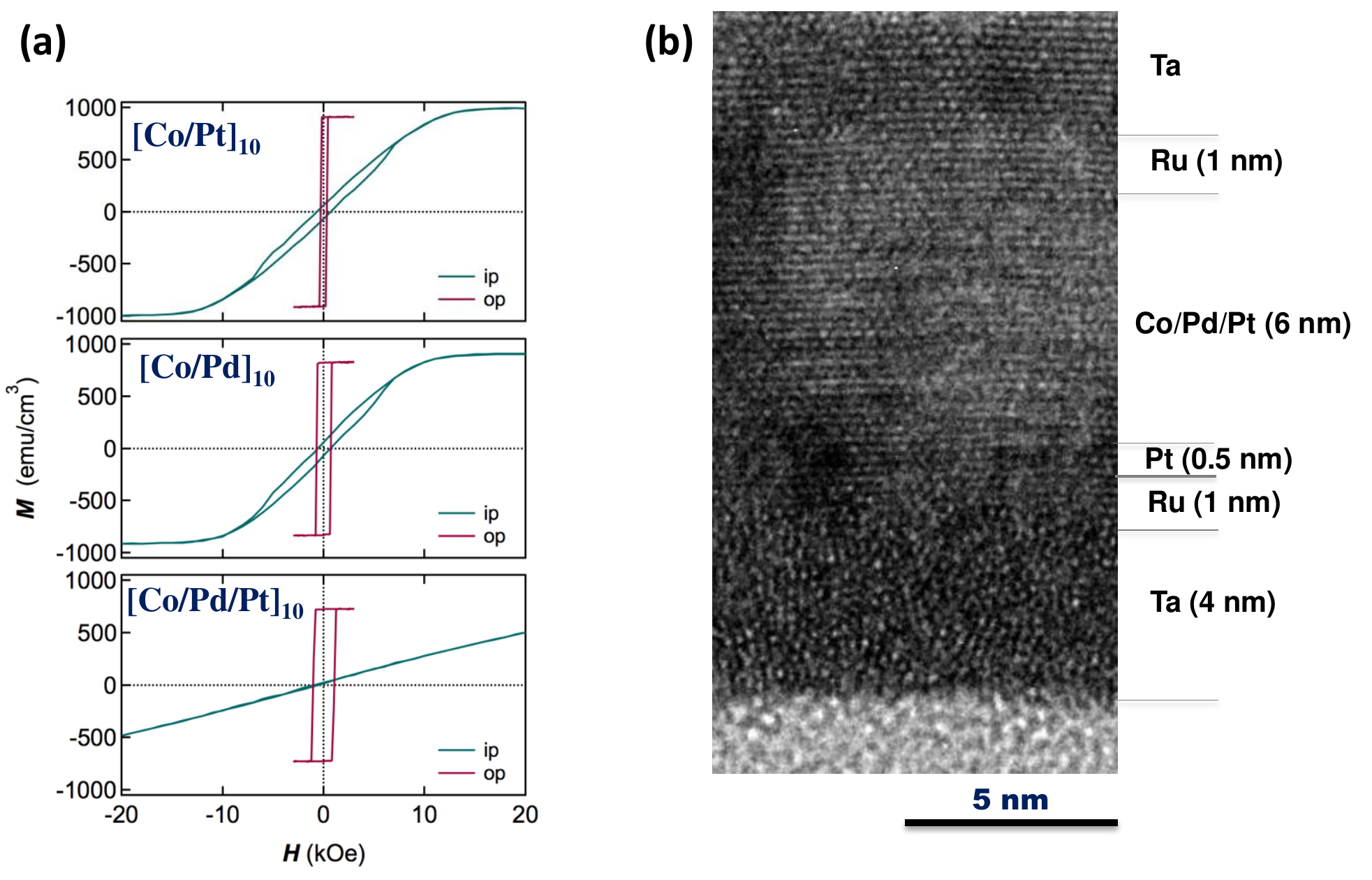}
\caption{(a) $M-H$ loops of [Co(0.2 nm)/Pt(0.2 nm)]$_{10}$, [Co(0.2 nm)/Pd(0.2 nm)]$_{10}$ and [Co(0.2 nm)/Pd(0.2 nm)/Pt(0.2 nm)]$_{10}$ films. The op and ip denote $M-H$ loops with out-of-plane and in-plane magnetic fields, respectively. (b) High-resolution TEM image for the full stack of the [Co(0.2 nm)/Pd(0.2 nm)/Pt(0.2 nm)]$_{10}$ film.}
\label{Expt}
\end{center}
\end{figure}

In order to study the effect of the symmetry of the stacking structure on the magnetic behavior, three variations of the multilayer structures were fabricated as follows: [Co(0.2 nm)/Pt (0.2 nm)]$_{10}$, [Co (0.2 nm)/Pd(0.2 nm)]$_{10}$ and [Co(0.2 nm)/Pd(0.2 nm)/Pt(0.2 nm)]$_{10}$. Our films have been fabricated with a sputtering apparatus (Canon-Anelva C-7100) on thermally oxidized Si substrates at room temperature. A Ta(4 nm)/Ru(1 nm)/Pt(0.5 nm) seed/buffer layer was first deposited on the substrate. Then a Co-based multilayer was grown by alternate deposition of Co, Pd and Pt at room temperature \cite{Yakushiji:2010}. 
Fig. \ref{Expt}a shows their $M-H$ (magnetization vs. magnetic field) loops measured at room temperature. All of them display typical perpendicularly magnetized properties as the sharp reversals in the out-of-plane (op) loop and the gradual saturation in the in-plane (ip) loop with a substantial perpendicular magnetic anisotropy field ($H_k$). The values of the effective PMA ($K_{\rm eff} = H_kM_s/2$) were estimated to be 5.7, 4.8 and 10.8 Merg/cm$^3$ derived from $H_k$ and the saturation magnetization ($M_s$), for [Co (0.2 nm)/Pt (0.2 nm)]$_{10}$, [Co (0.2 nm)/Pd (0.2 nm)]$_{10}$ and [Co (0.2 nm)/ Pd (0.2 nm)/Pt (0.2 nm)]$_{10}$, respectively. It is obvious that the asymmetric stacking, the Co/Pd/Pt-multilayer, exhibited twice larger $K_{\rm eff}$ than the symmetric stackings, the Co/Pt- and Co/Pd-multilayers. Fig. \ref{Expt}b shows the cross-sectional high-resolution transmission electron microscopy image for the full stack of the [Co(0.2 nm)/Pd(0.2 nm)/Pt(0.2 nm)]$_{10}$ multilayer. It suggests that an fcc(111) oriented Co/Pd/Pt-multilayer part was formed on the hcp-c-plane oriented Ru buffer layer with a flat interface. Larger $K_{\rm eff}$ values of the asymmetric samples with respect to those of the symmetric ones are consistently observed in various samples with different thicknesses, see Supplemental Material \cite{Supple}.

We now turn to our first-principles calculations in the basis of  Density Functional Theory (DFT) approach to CoPt$_2$, CoPd$_2$, and CoPtPd model systems in order to understand the origin of this behavior (see Supplemental Material \cite{Supple} for the detail of calculations). The structures resemble that of nonmagnetic noncentrosymmetric semiconductor BiTeI \cite{Ishizaka:2011}, which exhibits bulk Rashba spin-momentum coupling. 
The calculated $E_{\rm MCA}$ per unit volume are summarized in the first row of Table \ref{tabSummary}. From the positive $E_{\rm MCA}$ it is clearly visible that all systems possess PMA. 
Secondly, while CoPt$_2$ and CoPd$_2$ bulk systems have comparable MCA energies of around 0.027$-$0.028 meV/{\AA}$^3$, we find stronger MCA energies of more than 0.040 meV/{\AA}$^{3}$ for the CoPtPd system, in qualitative agreement with our experiment. 
If MCA arose entirely from each individual interface, the expected MCA for CoPtPd would be $0.028/2+0.027/2=0.0275$ meV/{\AA}$^{3}$ considering that there are two types of interfaces, Co/Pt and Co/Pd. This value is smaller than the calculated value of 0.043  meV/{\AA}$^{3}$ (see Table \ref{tabSummary}) by 0.0155 meV/{\AA}$^{3}$, which is sizable. Other contribution apart from the interfacial effect should therefore play a role in the larger MCA in CoPtPd system compared to the other cases, and it may be related to one of the following scenarios: 
(i) the difference in the number of valence electrons in all unit cells \cite{Daalderop:1992}; (ii) the difference in the SOC strength \cite{Stohr:1999}; (iii) the anisotropy of orbital magnetic moment, as proposed by Bruno \cite{Bruno:1989}; (iv) the different orbital hybridization that occurs in the considered systems \cite{Weller:1994}; and/or finally (v) the structural inversion symmetry breaking along $z$ \cite{Barnes:2014,Kim:2016} that leads to the bulk Rashba-type splitting \cite{Ishizaka:2011} in the CoPtPd case, compared to the other two systems.

The trivial scenarios (i) and (ii) can immediately be ruled out, as discussed in the Supplemental Material \cite{Supple}, due to the same total number of valence electrons in the unit cell, which is 29, and the presence of Pd in CoPtPd which has larger PMA than CoPt$_2$. Additionally, the anisotropy of the orbital magnetic moment, induced as a direct consequence of SOC, can be considered by defining $\mu_{\rm orb}^{\rm anis}=\mu_{\rm orb}^{\rm ip}-\mu_{\rm orb}^{\rm op}$. From our fully self-consistent SOC calculations, we obtain $\mu_{\rm orb}^{\rm anis}$ to be $-0.025$ $\mu_B$ for CoPt$_2$, which is larger than $-0.023$ $\mu_B$ for CoPtPd. Therefore, a larger MCA does not trivially correspond to a larger orbital moment anisotropy in the CoPtPd system here, thus making scenario (iii) inapplicable. Indeed, it has been pointed out that an extra care should be taken when considering the Bruno model for systems with strong spin-orbit coupling such as those containing $5d$ transition metal elements \cite{Andersson:2007}. 

To assess the relevance of scenario (iv), we calculated the Densities of States (DOSs) of these systems, as shown in Fig. S1 of the Supplemental Material \cite{Supple}. 
The calculated magnetic moments in all systems are found to be more than 2.0 $\mu_B$ for Co and around 0.3 $\mu_B$ for both Pt and Pd, hence the MCA should be driven by Co. We additionally calculate the relative contribution of each atomic layer to the PMA, by artificially switching on and off the SOC of the atoms. 
When the SOC of Co is switched off, in all considered cases the MCA vanish, 
confirming the crucial role of Co moment to drive the MCA. 
In Table \ref{tabSummary}, however, both CoPt$_2$ and CoPd$_2$ systems give comparable perpendicular MCA despite the nonnegligible difference in the Co$-d$ bandwidth in both cases. This observation suggests that the role of Co$-$Pt and Co$-$Pd orbital hybridizations in driving the PMA, as implied by the scenario (iv), is not significant. 
The remaining scenario which might explain the origin of the enhanced MCA in CoPtPd is therefore the structural inversion symmetry breaking along $z$ due to the presence of both Pt $and$ Pd layers sandwiching the Co layer, as suggested by scenario (v).

\begin{figure}
\begin{center}
\includegraphics[width=1.0\columnwidth]{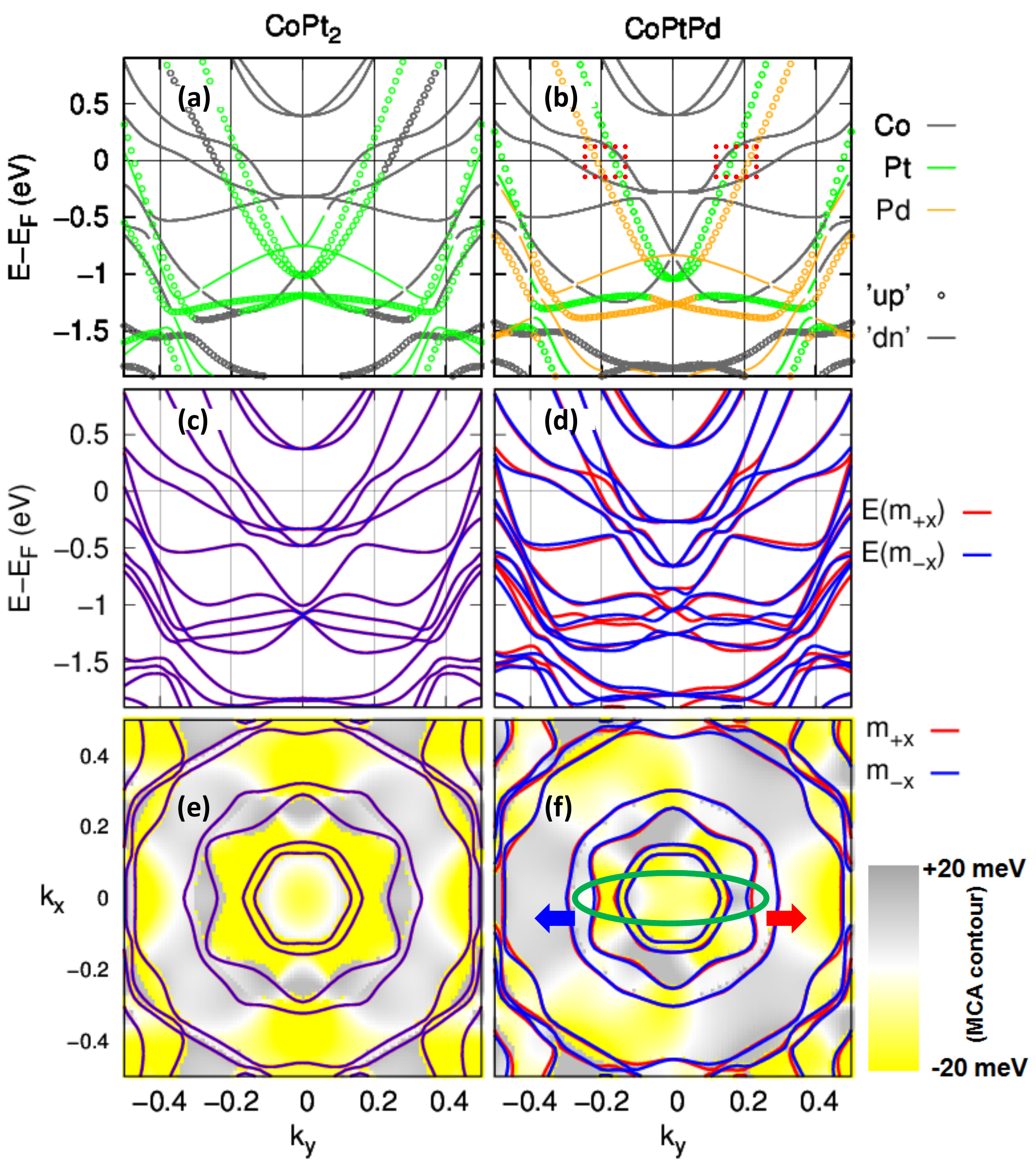}
\caption{Band structure along the $k_x=k_z=0$, i.e. along the (0,$-\frac{1}{2}$,0)$\rightarrow$(0,$\frac{1}{2}$,0) path, without spin-orbit coupling, (a) and (b), and with SOC, (c) and (d); and the two dimensional ($k_x,k_y$) in-plane Fermi surface, i.e. at $k_z=0$, together with the contour map of $k-$dependent MCA energy within this surface, (e) and (f). Left and right panels show the plots for CoPt$_2$ and CoPtPd, respectively. The m$_{\rm +x}$, m$_{\rm -x}$, and m$_{\rm z}$ labels indicate the magnetization along respectively the $+x$, $-x$, and $z$ directions. Likewise, E(m$_{\alpha}$) indicates the energy of $\alpha$ direction-imposed magnetization.}
\label{map2D}
\end{center}
\end{figure}

As in the case of nonmagnetic systems \cite{Bychkov:1984,Picozzi:2014,Manchon:2015}, the broken inversion symmetry in magnetic materials also results in the band splitting in the presence of SOC, as described by the Rashba model. This effect has previously been demonstrated in Gd(001) magnetic surfaces \cite{Krupin:2005}. We note however that the Rashba effect manifests differently in the nonmagnetic and magnetic systems, as can be summarized in the following \cite{Krupin:2005}: In nonmagnetic systems, the presence of Rashba interaction splits the otherwise degenerate up- and down-spin states. In ferromagnetic systems, the degeneracy of the up- and down-spins are already lifted, and 
Rashba interaction enhances/reduces the splitting depending on the sign of $\vec{k}\cdot(\hat{z}\times\vec{m})$. The resulting splitting is thus asymmetric with respect to $\vec{k}$ and the sign of asymmetry gets reversed when $\vec{m}$ is reversed. The asymmetric splitting is most pronounced along the $\vec{k}$ direction parallel to $\hat{z}\times\vec{m}$.

Therefore, in order to consider the role of structural asymmetry in the present model systems, we first visualize the spin splitting by following the prescription introduced in previous works \cite{Krupin:2005, Park:2013, Grytsyuk:2016}. 
We choose two different directions of in-plane magnetization, i.e. along $+x$ and $-x$ directions. 
The results are presented in Fig. \ref{map2D} for the CoPt$_2$ system representing the symmetric case and CoPtPd in which the inversion symmetry is broken. 
The band structures are presented in Figs. \ref{map2D}a-\ref{map2D}d along the ($k_x,k_y,k_z$)=(0,$-\frac{1}{2}$,0)$\rightarrow$(0,$\frac{1}{2}$,0) path. Near the Fermi level, the band structures are mainly dominated by the Co minority-spin states, as shown Figs. \ref{map2D}a and \ref{map2D}b without SOC. The majority-spin bands, on the other hands, are largely dispersive, and dominated by the Pt and Pd bands. 
When the SOC is switched on, the characteristic Rashba-type splitting for ferromagnetic system \cite{Krupin:2005} emerges in the asymmetric CoPtPd case (Fig. \ref{map2D}d), but is absent in the band structure of CoPt$_2$ case (Fig. \ref{map2D}c).

The Rashba-type splitting also manifests further in the distortion of Fermi contour (Figs. \ref{map2D}e and \ref{map2D}f). The Fermi surface of CoPt$_2$ does not show any shift in the $(k_x,k_y)$ plane for both in-plane magnetization directions. On the other hands, the Fermi surface of CoPtPd is shifted towards positive $k_y$ direction when the magnetization is oriented along $+x$ direction, as indicated by the red arrow in Fig. \ref{map2D}f. Additionally, switching the magnetization direction along the $-x$ direction, which is an opposite direction, gives a mirror symmetric Fermi contour along the $k_y$ \cite{Grytsyuk:2016}, i.e.  %
\begin{equation}
\label{k-splitting}
E(m_{+x},+k_y)=E(m_{-x},-k_y),
\end{equation}
as indicated by the blue arrow in Fig. \ref{map2D}f. Such mirror symmetry is disappearing along $k_x$ for this particular magnetization direction, as pointed out by Grytsyuk, \textit{et al.} \cite{Grytsyuk:2016}. 
The two-dimensional contour map of $E_{\rm MCA}$ within the ($k_x,k_y$) plane is fully symmetric along $k_y$ for CoPt$_2$, in contrast to the nonsymmetric CoPtPd case. 

We can proceed further by defining the $E_{\rm MCA}$ as the $E(m_{+x})-E(m_z)$ or $E(m_{-x})-E(m_z)$. Both definitions give the same total MCA energy as reported in Table \ref{tabSummary} when the evaluation is done within the whole Brillouin zone. 
However, the $k-$dependent $E_{\rm MCA}$ can provide an additional insight, since it can be virtually decomposed into $E_{\rm MCA}^{+k_y}$ and $E_{\rm MCA}^{-k_y}$, in which $E_{\rm MCA}^{\pm k_y}=E(m_x,\pm k_y)-E(m_z,\pm k_y)$. By integrating the $E_{\rm MCA}^{\pm k_y}$ within half of the Brillouin zone, i.e. within all $(k_x,k_z)$ space for each $+k_y$ and $-k_y$, respectively, we obtain the results as reported in Table \ref{tabSummary}. In this regard, one can define an even contribution
\begin{equation}
\label{even}
E_{\rm MCA}^{\rm even}=E_{\rm MCA}^{+k_y}+E_{\rm MCA}^{-k_y},
\end{equation}
which is exactly the $E_{\rm MCA}$ within the whole Brillouin zone. The estimation of this contribution to each model system is also shown in Table \ref{tabSummary}. The small difference that occurs between $E_{\rm MCA}$ (SCF) and $E_{\rm MCA}^{\rm even}$ clearly comes from the different SOC treatment in both cases, since the evaluation of $E_{\rm MCA}^{+k_y}$ and $E_{\rm MCA}^{-k_y}$ cannot be done self-consistently. Interestingly, another quantity,  
\begin{equation}
\label{odd}
E_{\rm MCA}^{\rm odd}=E_{\rm MCA}^{+k_y}-E_{\rm MCA}^{-k_y},
\end{equation}
can also be defined. 
This quantity provides an estimation on the shift of the Fermi surface, thus the estimation of the degree of inversion symmetry breaking. As summarized in the last row of Table \ref{tabSummary}, $E_{\rm MCA}^{\rm odd}$ is zero for both symmetric CoPt$_2$ and CoPd$_2$. 
The $E_{\rm MCA}^{\rm odd}$ in CoPtPd is on the other hand very large, implying that the system is highly nonsymmetric. 

To gain an additional insight on the role of the broken inversion symmetry, one may use a simple argument in which the MCA energy is expressed as $E_{\rm MCA}=\frac{\lambda}{\varepsilon_{\rm uo}}$ in the second-order perturbation theory, where $\varepsilon_{\rm uo}$ denotes the energy gap between the occupied and unoccupied states, while the $\lambda$ contains the spin-orbit interaction between these states and depends on the SOC coupling constant. The $E_{\rm MCA}^{+k_y}$ and $E_{\rm MCA}^{-k_y}$ can be given by $E_{\rm MCA}^{+k_y}=\frac{1}{2}\frac{\lambda}{\varepsilon_{\rm uo}-\Delta\varepsilon}$ and $E_{\rm MCA}^{-k_y}=\frac{1}{2}\frac{\lambda}{\varepsilon_{\rm uo}+\Delta\varepsilon}$, in which $\pm\Delta\varepsilon$ denotes the widening or narrowing of the energy gap due to the departure from the symmetric band structure, and hence is absent ($\Delta\varepsilon=0$) in the symmetric cases (Fig. \ref{map2D}c). The $\pm\Delta\varepsilon$ is therefore only present in the asymmetric systems, and in the case of CoPtPd (Fig. \ref{map2D}d), $\pm\Delta\varepsilon$ describes for instance the gap widening and narrowing at $k_y=-0.2$ and $k_y=0.2$. The total contribution is nothing but the $E_{\rm MCA}^{\rm even}$ in Eq. \eqref{even} as summarized in Table \ref{tabSummary}. However, in the specific cases, where the broken inversion symmetry is present ($\Delta\varepsilon\neq0$), the total MCA energy will be given by
\begin{equation}
E_{\rm MCA}^{\rm even}\approx\left(\frac{\lambda}{\varepsilon_{\rm uo}}\right)\frac{1}{1-(\frac{\Delta\varepsilon}{\varepsilon_{\rm uo}})^2}\cdot
\label{MCA_asym}
\end{equation}
Since $(\frac{\Delta\varepsilon}{\varepsilon_0})^2$ is always positive, Eq. \eqref{MCA_asym} indicates that the presence of $\Delta\varepsilon$
due to the asymmetry will lead to an increase of MCA. In other words, the enhancement of $E_{\rm MCA}^{\rm even}$ due to the inversion symmetry breaking is given by $\Delta E_{\rm MCA}^{\rm even}=\left(\frac{\Delta\varepsilon}{\varepsilon_{\rm uo}}\right)^2 E_{\rm MCA}^{\rm even}$, showing that such modification indeed occurs only in asymmetric systems, thus capturing the enhancement of MCA energy due to the spin-momentum coupling, as proposed recently \cite{Barnes:2014,Kim:2016}. 

At this stage, it is appealing to consider the origin of Rashba-type splitting in the CoPtPd. Returning to the band structure of CoPtPd (with SOC, Fig. \ref{map2D}d), the splitting can be seen for instance near the Fermi level around $k_y=\pm 0.2$ and at the energy of $-1$ eV around the $\Gamma$ point. Furthermore, this splitting is likely to be induced by the Pt and Pd states. Around the Fermi level, for example, the Rashba-type splitting at around $k_y=\pm 0.2$ is clearly dominated by the states indicated by the red broken rectangle in Fig. \ref{map2D}b. We further switched off the SOC of Co and we obtained an $E_{\rm MCA}^{\rm odd}$ of 0.193 meV/{\AA}$^{3}$, resembling closely the $E_{\rm MCA}^{\rm odd}$ in Table \ref{tabSummary}. Additionally, $E_{\rm MCA}^{\rm even}$ vanishes, showing the significance of the SOC of Co for the MCA. On the other hands, when only the SOC of Co is maintained while switching off the SOC of both Pt and Pd, the $E_{\rm MCA}^{\rm odd}$ diminishes down to $-0.010$ meV/{\AA}$^{3}$, confirming the crucial role of the SOC of Pt and Pd to drive the Rashba-type splitting. Interestingly, the $E_{\rm MCA}^{\rm even}$ also becomes practically zero in the latter case, implying that despite Co being the magnetic moment carrier, the SOC of Co alone is not sufficient to induce large MCA.

In real materials, interlayer mixing is likely to occur during the growth process. Such mixing can influence the MCA, especially for multilayer systems with small monolayer thickness as in our cases. We have therefore performed the calculation to estimate the effect of the intermixing (see Supplemental Material \cite{Supple}). We found that although the calculated MCA is quantitatively altered by the intermixing, the perpendicular MCA of CoPtPd is consistently larger than that of CoPt$_2$ case, indicating that the effect of inversion symmetry breaking on MCA is still effective in multilayers with interlayer mixing, although detailed effects of the structural disordering to the MCA require a further study.
 
Finally we note the similarity between the structure of CoPtPd system in the present work and that of the nonmagnetic noncentrosymmetric semiconductor BiTeI \cite{Ishizaka:2011}. The crystal structure of BiTeI consists of triangular network of single layers of Bi, Te, and I. In this system, a giant bulk Rashba-type splitting has also been recently observed \cite{Ishizaka:2011}. Our work thus highlights the presence of bulk Rashba-type effect due to the broken inversion symmetry in magnetic materials, as well as a possible direct consequence in terms of the enhanced MCA.

In summary, from our combined first-principles calculation and experimental results, we provide for the first time an evidence for the asymmetric-structure-driven enhancement of PMA in the transition metal multilayers. In agreement with previous prediction on the MCA modulation due to the change in the Rashba parameter \cite{Kim:2016}, we show that the breaking of inversion symmetry does not only lead to a modification, but also to an enhancement, to the PMA. While the PMA is realized through Co which carries the magnetic moments, the breaking of inversion symmetry can significantly enhance the perpendicular MCA strength due to the presence of Pt and Pd with strong SOC. This work may provide a guideline on the design of materials with strong PMA, as well as a suggestion to the investigation of possible enhancement of other SOC-related properties, such as the anomalous and spin Hall effect, due to the broken structural inversion symmetry.

\begin{acknowledgments}
A.-M. P. was supported by the Japan Society for the Promotion of Science (JSPS) KAKENHI Grant No. 15H05702. H.-W. L. acknowledges financial support from the National Research Foundation of Korea (NRF, Grant No. 2018R1A5A6075964). K.-J. L. and K.-W. K. acknowledge the KIST Institutional Program (Projects No. 2V05750 and No. 2E29410). K.-W. K. also acknowledges financial support from the German Research Foundation (DFG) (No. SI 1720/2-1). Work was also in part supported by JSPS KAKENHI Grant No. 16K05415, the Cooperative Research Program of Network Joint Research Center for Materials and Devices, and Center for Spintronics Research Network (CSRN), Osaka University. Computations were performed at Research Institute for Information Technology, Kyushu University.
\end{acknowledgments}

\bibliographystyle{apsrev}

\end{spacing}
\end{document}


\title{Supplemental Material for ``Enhanced perpendicular magnetocrystalline anisotropy energy in an artificial magnetic material with bulk spin-momentum coupling''}

\author{Abdul-Muizz Pradipto}
\email[Electronic address: ]{a.m.t.pradipto@gmail.com}
\affiliation{Institute for Chemical Research, Kyoto University, Uji, Kyoto 611-0011, Japan}
\affiliation{Department of Physics Engineering, Mie University, Tsu, Mie 514-8507, Japan}

\author{Kay Yakushiji}
\affiliation{National Institute of Advanced Industrial Science and Technology (AIST), Tsukuba, Ibaraki 305-8568, Japan}

\author{Woo Seung Ham}
\affiliation{Institute for Chemical Research, Kyoto University, Uji, Kyoto 611-0011, Japan}

\author{Sanghoon Kim}
\affiliation{Institute for Chemical Research, Kyoto University, Uji, Kyoto 611-0011, Japan}

\author{Yoichi Shiota}
\affiliation{Institute for Chemical Research, Kyoto University, Uji, Kyoto 611-0011, Japan}

\author{Takahiro Moriyama}
\affiliation{Institute for Chemical Research, Kyoto University, Uji, Kyoto 611-0011, Japan}

\author{Kyoung-Whan Kim}
\affiliation{Institute of Physics, Johannes Gutenberg University Mainz, 55099 Mainz, Germany}
\affiliation{Center for Spintronics, Korea Institute of Science and Technology, Seoul 02792, Korea}

\author{Hyun-Woo Lee}
\affiliation{Department of Physics, Pohang University of Science and Technology, Pohang 37673, Korea}

\author{Kohji Nakamura} 
\affiliation{Department of Physics Engineering, Mie University, Tsu, Mie 514-8507, Japan}

\author{Kyung-Jin Lee}
\affiliation{Department of Materials Science and Engineering, Korea University, Seoul 02841, Korea}
\affiliation{KU-KIST Graduate School of Converging Science and Technology, Korea University, Seoul 02841, Korea}

\author{Teruo Ono}
\affiliation{Institute for Chemical Research, Kyoto University, Uji, Kyoto 611-0011, Japan}


\begin{spacing}{1.2}

\maketitle

\section{Layer thickness dependence of effective magnetocrystalline anisotropy constants}

\begin{figure*}
\begin{center}
\includegraphics[width=1.0\columnwidth]{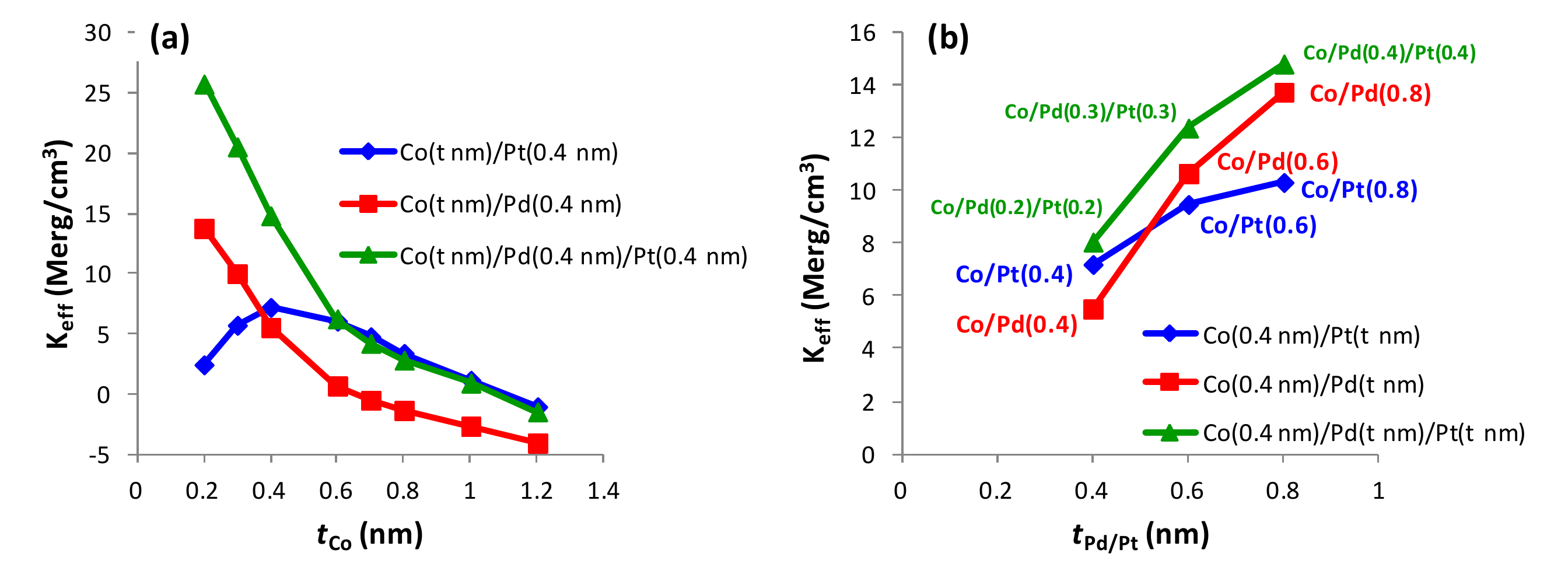}
\caption{(a) Co thickness dependence, with fixed thickness of heavy metal layers of 0.4 nm, and (b) heavy metal thickness dependence, with fixed Co thickness of 0.4 nm, of the magnetocrystalline anisotropy (MCA) of [Co/Pt]$_n$, [Co/Pd]$_n$, and [Co/Pd/Pt]$_n$ superlattices.}
\label{thickness}
\end{center}
\end{figure*}

A series of samples representing the repetitive [Co/Pt]$_n$, [Co/Pd]$_n$, and [Co/Pd/Pt]$_n$ multilayer structures have been fabricated with various thicknesses according to the procedure described previously \cite{Yakushiji:2010}. In Fig. \ref{thickness}a, we show the dependence of effective magnetocrystalline anisotropy (MCA) values, i.e. $K_{\rm eff}=H_kM_s/2$, as functions of Co layer thickness. While the $K_{\rm eff}$ of [Co/Pt]$_n$ multilayers shows an optimum value at Co thickness of around 0.4 nm, the $K_{\rm eff}$ values of [Co/Pd]$_n$ and [Co/Pd/Pt]$_n$ multilayers tend to keep increasing with the decreasing Co thickness, hence no optimum Co thickness have been implied for [Co/Pd]$_n$ and [Co/Pd/Pt]$_n$ multilayes. We therefore consider 0.4 nm to be the optimum thickness of Co layer and fix the Co thickness to 0.4 nm to see further the behavior of $K_{\rm eff}$ on the various heavy metal layer thicknesses, as summarized in Fig. \ref{thickness}b. 

The different behaviors of $K_{\rm eff}$ in [Co/Pt]$_n$ and [Co/Pd]$_n$ multilayer with the decreasing Co thicknesses as shown in Fig. \ref{thickness}a are in accordance with previous works on the sputtered multilayers \cite{Canedy:2000,Kato:2012}. In [Co/Pd]$_n$ multilayer, the absence of peak in the $K_{\rm eff}$ at various Co thicknesses was also shown \cite{Kato:2012}. On the other hands, an optimum Co thickness of about $4-6$ {\AA} in Co/Pt multilayer has also been reported \cite{Canedy:2000}, which is very similar to our results, despite the absence of multilayer repetition. It turns out that the behavior of [Co/Pd/Pt]$_n$ is resembling [Co/Pd]$_n$ multilayer in this regard. 

In Fig. \ref{thickness}b, all $K_{\rm eff}$ values are visibly enhanced with the increasing heavy metal thicknesses. However, if one compares the $K_{\rm eff}$ of samples with the same thickness of heavy metals, e.g. [Co(0.4)/Pt(0.4)]$_n$, [Co(0.4)/Pd(0.4)]$_n$, and [Co(0.4)/Pd(0.2)/Pt(0.2)]$_n$ multilayers, an important behavior can be discerned. As discussed in the main text, if only the interfacial effect is crucial to determine the MCA, the expected value of $K_{\rm eff}$ of [Co/Pd/Pt]$_n$ would be the average of those of [Co/Pd]$_n$ and [Co/Pt]$_n$. However, in samples with same heavy metal thicknesses, the MCA constant of the asymmetric system [Co/Pd/Pt]$_n$ is always larger than those of the other systems, even when an optimum Co thickness in [Co/Pt]$_n$ has been assumed, with visibly larger $K_{\rm eff}$ than the average values of [Co/Pd]$_n$ and [Co/Pt]$_n$. A consistently larger perpendicular MCA in the [Co/Pd/Pt]$_n$ case should therefore support the crucial role of bulk spin-orbit coupling effect in driving the enhanced MCA in the asymmetric multilayer.

\section{First-principles calculations}

\begin{figure*}
\begin{center}
\includegraphics[width=1.0\columnwidth]{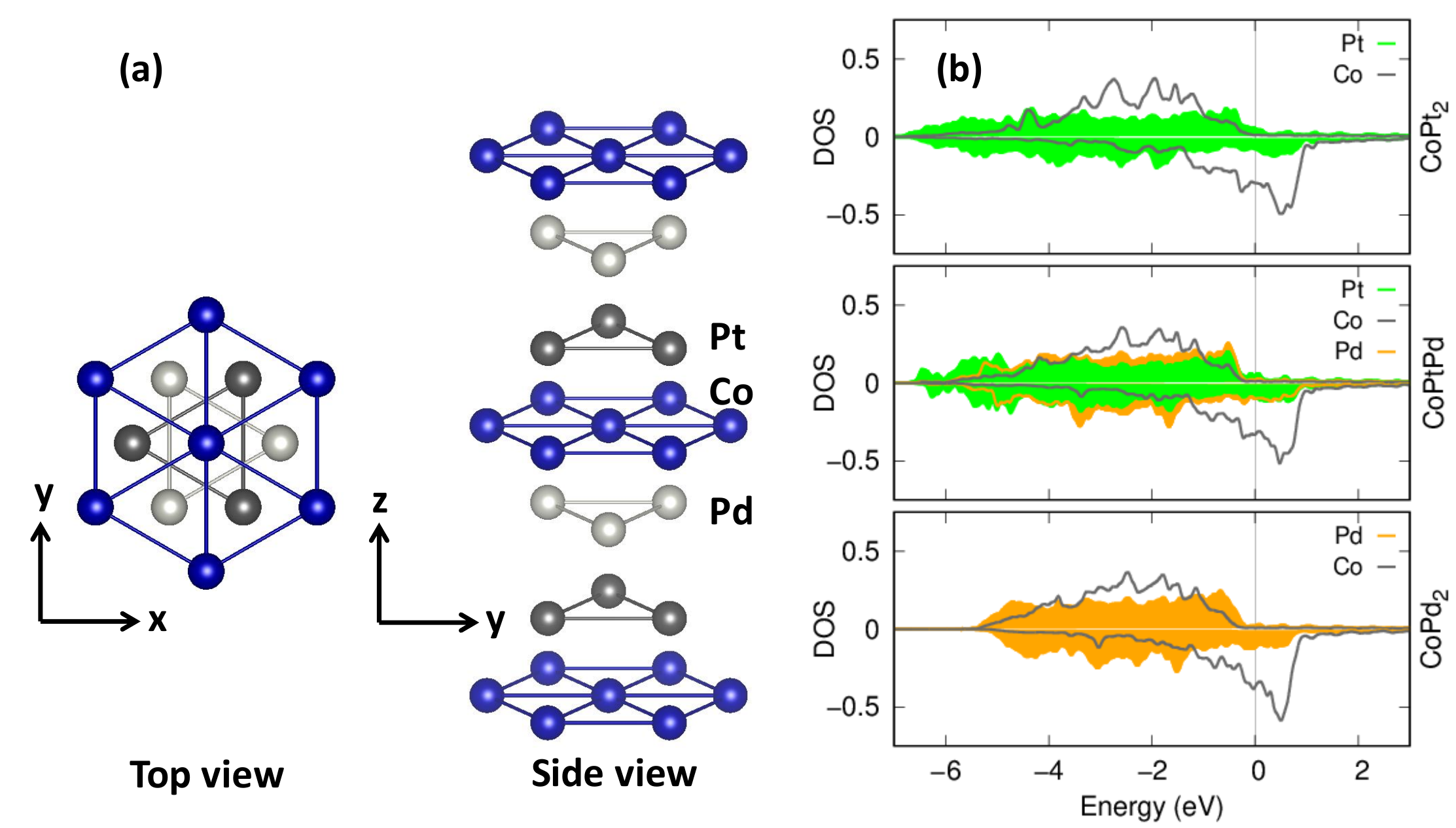}
\caption{(a) The model crystal structure of CoPtPd multilayer system. For CoPt$_2$ (CoPd$_2$) model system, the Pd (Pt) layer is replaced by another Pt (Pd) atomic layer. (b) The Density of States (DOS) of CoPt$_2$, CoPtPd, and CoPd$_2$ systems.}
\label{DOS}
\end{center}
\end{figure*}

Our calculations employ the Generalized Gradient Approximation (GGA) \cite{Perdew:1996} functional as implemented in the Full-Potential Linearized Augmented Plane Wave (FLAPW) code \cite{Wimmer:1981,Weinert:1982,Nakamura:2003}. 
The in-plane lattice constant has been fixed to the optimized value of bulk Pt, but the out-of-plane atomic positions and lattice constants, i.e. normal to the layer plane, have been relaxed for all systems. 
The MCA energy $E_{\rm MCA}$ has been defined in the calculations as $E_{\rm MCA}=E_{\rm ip}-E_{\rm op}$, where $E_{\rm ip}$ ($E_{\rm op}$) refers to the total energy of the in- (out-of-)plane magnetization, implying that positive $E_{\rm MCA}$ indicates the preferred out-of-plane magnetization. The SOC has been included self-consistently, so that full relaxation of the charge and spin densitites is allowed upon the treatment of SOC. The MCA energies obtained in this self-consistent field (SCF) manner are indicated by $E_{\rm MCA}$ (SCF) in Table 1 of the main text. Our convergence test calculations show that the MCA energies do not vary larger than 10$^{-4}$ meV/{\AA}$^3$ after $75\times 75\times 45$ $k-$point mesh which gives a total of 253125 $k-$points in the Brillouin zone. In order to model the multilayers, three different bulk systems have been chosen in the calculations so that the unit cells contain three layers of Pt/Co/Pt (referred to hereafter as CoPt$_2$ model system), Pd/Co/Pd (referred as CoPd$_2$), and Pt/Co/Pd (referred as CoPtPd) stacked in an fcc(111) pattern, such that the Co atoms are sandwiched between 2 Pt layers, 2 Pd layers, or between a Pt layer and a Pd layer, respectively (see Fig. \ref{DOS}a). 

Scenario (i), as discussed in the main text, for explaining the different MCA on the basis of the difference in electron number \cite{Daalderop:1992} can immediately be ruled out since all systems have the same total number of valence electrons in the unit cell, which is 29. Furthermore, in order to discuss scenario (ii) which is based on the SOC strength \cite{Stohr:1999}, we notice that in CoPt$_2$ unit cell there are two Pt, i.e. $5d$, atoms with a SOC strength of about 320 meV. On the other hands, CoPtPd contains Pd, a $4d$ element, whose SOC strength is around 104 meV, i.e. three times weaker than the $5d$ atom. The other system, CoPd$_2$, does not even contain Pt in the unit cell. However, both CoPt$_2$ and CoPd$_2$ have equally strong MCA and CoPtPd has decidedly the largest perpendicular MCA among these systems, hence a na\"ive view just by considering the SOC strength as in scenario (ii) does not hold in this case. The Density of States of these systems, relevant for the discussion of scenario (iv) based on the different orbital hybridization in the main text, are shown in  Fig. \ref{DOS}b. The majority- and minority-spin features of the DOSs suggest that the magnetic moments are mainly carried out by Co. 

\section{Structural disordering}

\begin{figure*}
\begin{center}
\includegraphics[width=1.0\columnwidth]{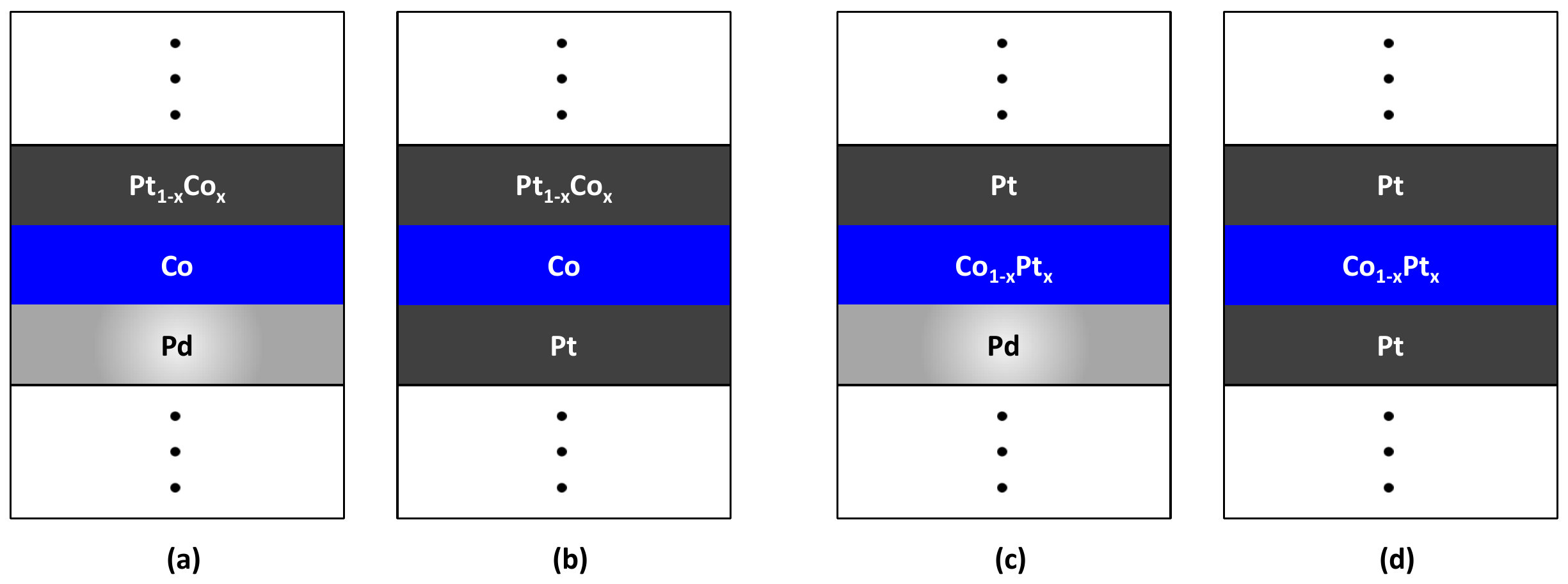}
\caption{The structural disordering models: (a) and (b) Pt$_{1-x}$Co$_x$ structures to model the cases where some Co amounts mix within the Pt monolayer of CoPtPd and CoPt$_2$ systems, respectively; (c) and (d) Co$_{1-x}$Pt$_x$ structures to describe the mixing of Pt in the Co monolayer of CoPtPd and CoPt$_2$ systems, respectively.}
\label{alloyingFig}
\end{center}
\end{figure*}

To consider the effects of structure disordering during the synthesis of the materials to the MCA, we generate supercells by doubling the unit cells of CoPtPd and CoPt$_2$ to represent respectively the asymmetric and the symmetric structures along both in-plane directions. We further introduce the in-plane alloying between Co and Pt, as depicted in Fig. \ref{alloyingFig}. Due to the increase of the system sizes, we only attempt the second-variational \cite{Li:1990,Wu:1999}, instead of the self-consistent, treatment of the SOC for the calculation of MCA. The results are summarized in Table \ref{alloyingTab}. As can be seen for the calculated MCA values of the ideal structures, i.e. without disorder, our second-variational SOC treatment has reproduced reasonably well the MCA energy obtained from the self-consistent treatment for the unit cell (Table 1 of the main text). 

When the layer intermixing is introduced, in all systems, the MCA is visibly affected by the composition of Co and Pt in the layer. 
One may notice that the introduction of certain amounts of Co composition to the Pt layer (i.e. the Pt$_{1-x}$Co$_x$ structures in Fig. \ref{alloyingFig}b) clearly destroys the inversion symmetry on the otherwise symmetric CoPt$_2$. Therefore, while the perpendicular MCA of CoPtPd is still larger than that of CoPt$_2$, discussing the difference of MCA in terms of inversion symmetry breaking would become irrelevant. On the other hand, the inversion symmetry may be maintained in the Co$_{1-x}$Pt$_x$ structures (Fig. \ref{alloyingFig}d), i.e. when some Pt amounts mix within the Co monolayer. When a Pt atom intermixes within the Co layer, a larger spin moment of about 0.39 $\mu_B$ is induced at the heavy metal atom, compared to 0.35 $\mu_B$ and 0.37 $\mu_B$ for the Pt atoms at the ideally ordered structures of CoPtPd and CoPt$_2$, respectively. In such cases, the calculated MCA is quantitatively altered by the intermixing, with a much more pronounced effect on the CoPt$_2$ case. 
One may also note that in the present calculations, the mixing of Co and Pt with $x=0.25$ is rather large. But even though it can significantly reduce the sample quality, a general trend is qualitatively maintained, i.e. the larger perpendicular MCA in CoPtPd than that in CoPt$_2$. Quantitatively, a modeling that requires huge unit cells with many configurations inducing different atoms and vacancies or a different theoretical framework \cite{Bansil:1991} may be desirable, which is beyond the scope of the present study.


\begin{table*}
\begin{center}
\caption{The MCA energy $E_{\rm MCA}$ per unit volume (in meV/{\AA}$^3$) obtained from the supercell calculations for the disordered structures, as depicted in Fig. \ref{alloyingFig}.}
\begin{tabular}{lcc*{1}{D{.}{.}{4}}@{}cc*{1}{D{.}{.}{5}}@{}}
\toprule[1pt]
System &&& \multicolumn{1}{c}{CoPtPd} &&& \multicolumn{1}{c}{CoPt$_2$} \\ 
 \midrule[0.5pt]
\rule{0pt}{2.5ex}Ideal structures &&& 0.042 &&& 0.030 \\
\rule{0pt}{2.5ex}Pt$_{0.75}$Co$_{0.25}$ (Figs. \ref{alloyingFig}a and \ref{alloyingFig}b)  &&& 0.027 &&& 0.024 \\
\rule{0pt}{2.5ex}Co$_{0.75}$Pt$_{0.25}$ (Figs. \ref{alloyingFig}c and \ref{alloyingFig}d) &&& 0.043 &&& -0.007 \\
\bottomrule[1pt]
\end{tabular}
\label{alloyingTab}
\end{center}
\end{table*}

%
%

\bibliographystyle{apsrev}


\end{spacing}